\documentclass[twocolumn]{aastex62}
\shorttitle{Dissecting a GRB fireball}
\shortauthors{Zhang et al.}
\begin{document}
\title{\bf Dissecting the Energy Budget of a Gamma-Ray Burst Fireball}

\author[0000-0002-9725-2524]{Bing Zhang}
\affiliation{Department of Physics and Astronomy, University of Nevada Las Vegas, Las Vegas, NV 89154, USA; zhang@physics.unlv.edu}

\author{Yu Wang}
\affiliation{ICRANet, Piazza della Repubblica 10, 65122 Pescara, Italy}
\affiliation{INAF -- Osservatorio Astronomico d'Abruzzo, Via M. Maggini snc, I-64100, Teramo, Italy}
\affiliation{Dip. di Fisica and ICRA, Sapienza Universita di Roma, Piazzale Aldo Moro 5, I-00185 Rome, Italy}

\author[0000-0002-1343-3089]{Liang Li}
\affiliation{ICRANet, Piazza della Repubblica 10, 65122 Pescara, Italy}
\affiliation{INAF -- Osservatorio Astronomico d'Abruzzo, Via M. Maggini snc, I-64100, Teramo, Italy}
\affiliation{Dip. di Fisica and ICRA, Sapienza Universita di Roma, Piazzale Aldo Moro 5, I-00185 Rome, Italy}


\begin{abstract}
The jet composition and radiative efficiency of GRBs are poorly constrained from the data. If the jet composition is matter-dominated (i.e. a fireball), the GRB prompt emission spectra would include a dominant thermal component originating from the fireball photosphere and a non-thermal component presumably originating from internal shocks whose radii are greater than the photosphere radius.  We propose a method to directly dissect the GRB fireball energy budget into three components and measure their values by combining the prompt emission and early afterglow data. The measured parameters include the initial dimensionless specific enthalpy density ($\eta$), bulk Lorentz factors at the photosphere radius ($\Gamma_{\rm ph}$) and before fireball deceleration ($\Gamma_0$), the amount of mass loading ($M$) and the GRB radiative efficiency ($\eta_\gamma$). All the parameters can be derived from the data for a GRB with a dominant thermal spectral component, a deceleration bump feature in the early afterglow lightcurve, and a measured redshift. The results only weakly depend on the density $n$ of the interstellar medium when the composition ${\cal Y}$ parameter (typically unity) is specified. 
\end{abstract}


\section{Introduction} \label{sec:intro}

The jet composition of the gamma-ray bursts (GRBs) has been subject to debate \citep{kumarzhang15,peer15,zhang18}. The GRB prompt emission spectra can in principle help to diagnose the jet composition: the existence of a bright thermal component would support a matter-dominated fireball \citep{meszarosrees00}, while the non-detection of such a component may suggest the dominance of a Poynting flux in the jet composition \citep{zhangpeer09}\footnote{A thermal component may still show up if the central engine magnetization parameter $\sigma_0$ is not extremely large so that $\sigma$ at the photosphere already drops to close to unity  \citep[e.g.][]{gaozhang15,Beniamini2017}.}. Broad band observations with GRB detectors, especially with the Gamma-ray Burst Monitor (GBM) and Large Area Telescope (LAT) on board the {\em Fermi} Gamma-Ray Space Telescope, have collected rich data, which suggest that the GRB jet composition is likely diverse. Whereas some GRBs (e.g. GRB 090902B, \citealt{abdo09c,ryde10,peer12b}, see \citealt{ryde05,ryde09,lil19} for systematic searches) are consistent with being fireballs, a good fraction of bursts are consistent with not having a thermal component \citep[e.g. GRBs 080916C, 130606B, and many others, ][]{abdo09a,zhang11,zhangbb16a,oganesyan17,ravasio19,burgess20}. ``Intermediate'' GRBs with a dominant non-thermal component and a sub-dominant thermal component have been discovered \citep[e.g. GRB 100724B, GRB 110721A and several others,][]{guiriec11,guiriec15,axelsson12}, which may be understood within the framework of ``hybrid'' jets, i.e. the composition is a mixture of a matter component and a Poynting-flux component \citep{gaozhang15,lil20}. Some bursts (e.g. GRB 160625B) displayed a significant change of jet composition among different emission episodes within the same GRB \citep{zhangbb18a,lil19b}, which may be consistent with some central engine models \citep[e.g.][]{metzger11}. Different jet compositions may imply different energy dissipation (shocks vs. magnetic reconnection) and radiation (quasi-thermal vs. synchrotron) mechanisms. 

Another interesting subject related to the GRB prompt emission mechanism is the radiative efficiency of a burst, which may be defined as \citep{lloydronningzhang04}
\begin{eqnarray}
    \eta_\gamma & \equiv & \frac{E_\gamma}{E_{\rm tot}} = \frac{E_\gamma}{E_\gamma+E_k} = \frac{L_\gamma}{L_{w,0}}, \label{eq:eta_gam} \end{eqnarray}
where $E_\gamma$, $E_k$ and $E_{\rm tot}$ are isotropic-equivalent gamma-ray energy, afterglow kinetic energy, and total energy, respectively, and $L_\gamma$ and $L_{w,0}$ are the isotropic-equivalent average gamma-ray luminosity and total wind luminosity at the central engine, respectively.  Considering beaming correction would lead to the same results, since all the energy/luminosity terms are multiplied by the same beaming factor $f_b$, which is not considered in the discussion below. The $E_\gamma$ value can be well measured from the data as long as the fluence is well measured and redshift is known. The $E_k$ term, on the other hand, is usually estimated from the afterglow data through modeling. Its value depends on many uncertain shock microphysics parameters, mostly $\epsilon_e$ \citep{freedman01}, but also $\epsilon_{\rm B}$ and electron spectral index $p$ as well \citep{zhang07a,wang15}. As a result, the derived GRB radiative efficiency has been subject to large uncertainties, ranging from below 10\% to more than 90\% \citep{zhang07a,wang15,Beniamini2015,Li2018}. 

The bulk Lorentz factor $\Gamma$ of a GRB, which is related to the kinetic energy of the outflow, has been estimated using various methods. The maximum photon energy of prompt emission may be used to set a lower limit on $\Gamma$ \citep[e.g.][]{baring97,lithwick01}. However, a precise measurement cannot be made since the maximum energy also depends on emission radius, which is not well constrained \citep{gupta08}\footnote{Most work made use of the variability timescale to estimate the emission radius, but the estimate is only relevant for the internal shock model but does not apply to photosphere \citep[e.g.][]{rees05} or magnetic dissipation \citep[e.g.][]{meszarosrees97b,Lyutikov2003,zhangyan11} models.}. Two other methods can give better estimates of $\Gamma$: The first makes use of the early afterglow lightcurve data. If a well-defined bump is identified in the early afterglow lightcurve, it can be interpreted as the fireball deceleration time. The Lorentz factor before deceleration (which we call $\Gamma_0$ in the rest of the paper) can be estimated \citep{rees92,meszarosrees93,saripiran99}, which depends on $E_k$ and the medium density parameter (i.e. $n$ for the constant medium model and $A_*$ for the wind model). Again $E_k$ needs to be estimated from the afterglow data or from the prompt emission data assuming an efficiency parameter. Alternatively, if a strong thermal component is measured from the GRB prompt emission spectrum, one can estimate the Lorentz factor at the photosphere radius (which we call $\Gamma_{\rm ph}$ in the rest of the paper) based on the standard fireball photosphere model \citep{peer07}. The GRB efficiency again needs to be assumed in order to perform the estimate. This simple method relies on the assumption of a matter dominated jet composition. For more general hybrid jet models, more complicated diagnoses are needed \citep{gaozhang15}. Observationally, the $\Gamma$ values derived from the afterglow deceleration method \citep{liang10,lv12,Ghirlanda2018} is somewhat smaller than those derived using other methods \citep{racusin09,peer15b}.

In this paper, we propose a new method to diagnose fireball parameters by combining the deceleration and photosphere methods. We show that with adequate observations, one can {\em measure} several fireball parameters related to the energy budgets. In particular, the efficiency parameter that has to be assumed in previous methods can be directly measured. The method is introduced in Section \ref{sec:method}. Some examples are presented in Section \ref{sec:example}. The results are summarized in Section \ref{sec:con} with some discussion.

\section{The method}\label{sec:method}

\subsection{Energy budget decomposition}\label{sec:energy}

\begin{figure*}[t]
\plotone{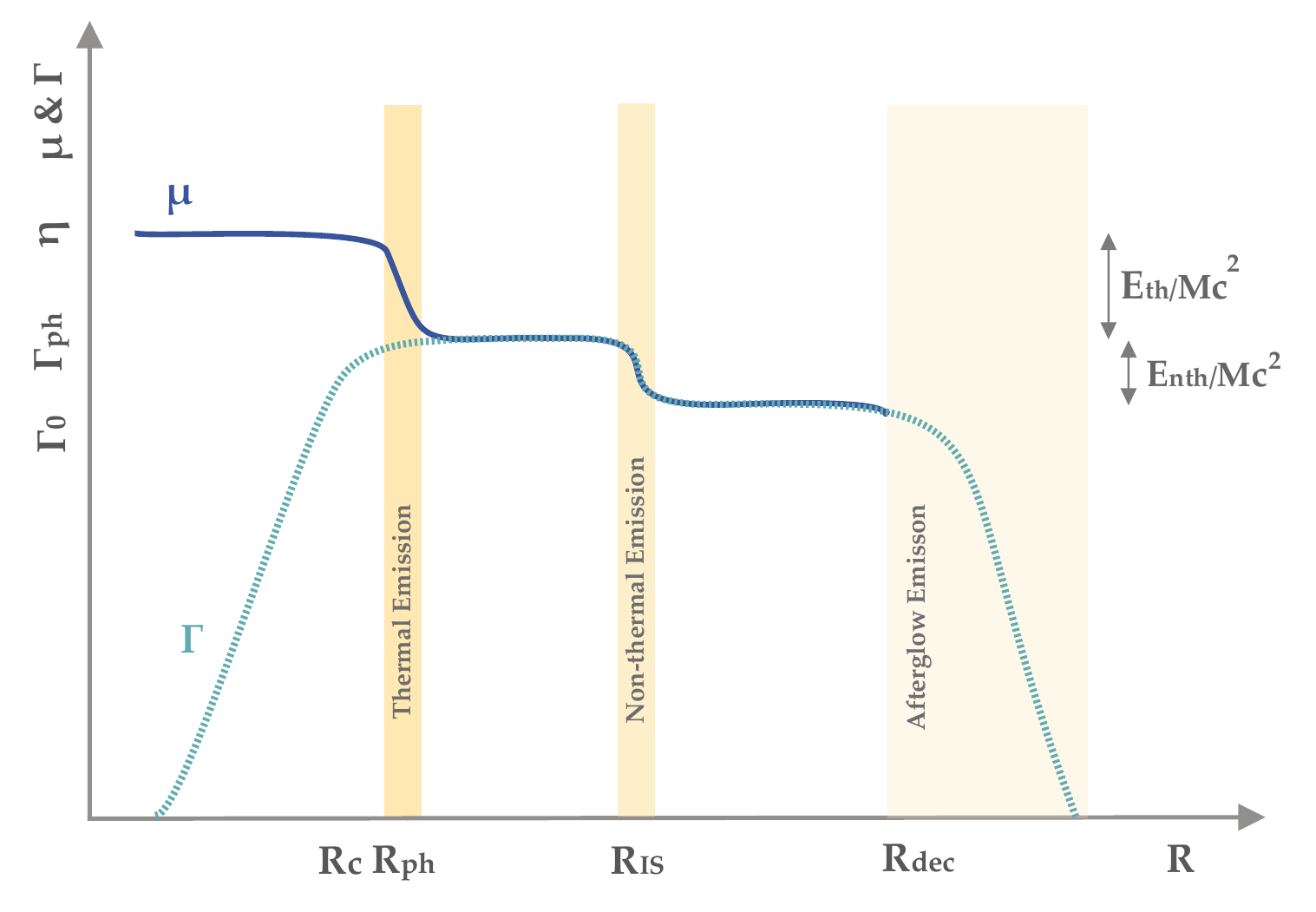}
\caption{An indicative description of the evolution of $\mu$ and $\Gamma$ in a GRB fireball. Both axes are in logarithmic scales. In reality, internal shocks may spread in a wide range of radii.}
\label{fig:cartoon}
\end{figure*}

Very generally, the effective energy per baryon at the central engine can be defined by the parameter
\begin{equation}
    \mu_0 \equiv \eta (1+\sigma_0) \simeq \eta,
\label{eq:mu0}
\end{equation}
where $\eta \equiv (n_{w,0} m_p c^2 + e_0 + p_0)/(n_{w,0} m_p c^2) = 1 + \hat\gamma e_0/(n_{w,0} m_p c^2)$, $\sigma_0$, $n_{w,0}$, $e_0$, $p_0$ are the dimensionless specific enthalpy density (also called dimensionless entropy in the literature, e.g. \citealt{meszarosrees00}), the magnetization parameter, number density, internal energy density, and pressure of the fireball wind at the central engine, respectively, and $\hat\gamma = 4/3$ is the adiabatic index for a relativistic fireball with $\eta  \gg 1$. The last approximation in Equation (\ref{eq:mu0}) applies to a pure fireball with $\sigma_0 \simeq 0$, which is the regime discussed in this paper. During the subsequent evolution of the fireball, the effective energy per baryon can be defined by
\begin{equation}
    \mu = \Gamma(R) \Theta(R), 
\end{equation}
which is conserved unless radiation is leaked out from the fireball. Here $\Gamma(R)$ is the bulk Lorentz factor of the fireball as a function of the radius $R$ from the central engine, and $\Theta (R) = 1 + \hat\gamma e(R)/[n_{w}(R) m_p c^2]$ is the dimensionless specific enthalpy density as a function of $R$. Figure \ref{fig:cartoon} shows a cartoon picture of the evolution of $\mu$ (only up to the deceleration radius $R_{\rm dec}$, beyond which it is no longer of interest) and $\Gamma$ (throughout the acceleration, coasting, dissipation, and deceleration phases) as a function of $R$. One can see that before the deceleration radius, the $\mu$ parameter undergoes two significant drops: The first drop occurs at the photosphere radius where a significant amount of thermal energy is released as thermal photons. The $\mu$ value drops from $\eta$ to $\Gamma_{\rm ph}$. The second drop occurs at the internal shock radii where significant dissipation of the fireball kinetic energy occurs and additional photon energy (in the form of synchrotron radiation) is released from the fireball. The $\mu$ value drops from $\Gamma_{\rm ph}$ to $\Gamma_{0}$ before entering the deceleration phase.

For a fireball with an isotropic equivalent total mass $M$, the initial, total energy of the fireball is
\begin{equation}
    E_{\rm tot} = \eta M c^2.
\end{equation}
The energy emitted in thermal emission from the photosphere is
\begin{equation}
    E_{\rm th} = (\eta-\Gamma_{\rm ph}) M c^2;
\label{eq:Eth}
\end{equation}
that emitted in non-thermal emission from internal shocks is
\begin{equation}
    E_{\rm nth} = (\Gamma_{\rm ph}-\Gamma_0) M c^2;
\label{eq:Enth}
\end{equation}
and the total emitted energy is 
\begin{equation}
    E_{\gamma} = E_{\rm th} + E_{\rm nth} = (\eta-\Gamma_0) M c^2.
\end{equation}
The kinetic energy left in the afterglow is 
\begin{equation}
    E_k = \Gamma_0 M c^2,
\end{equation}
so that the radiative efficiency  (\ref{eq:eta_gam}) becomes
\begin{equation}
    \eta_\gamma =  \frac{\eta-\Gamma_0}{\eta}. 
\label{eq:eta_gam2}
\end{equation}

\subsection{Prompt emission constraint}\label{sec:prompt}

The fireball initially undergoes a rapid acceleration with $\Gamma \propto R$ due to the internal pressure of the fireball \citep{meszaros93,piran93,kobayashi99}. It coasts at a radius $R_c = \Gamma_c R_0$ at which acceleration essentially stops, where $R_0$ is the initial radius of the fireball, and $\Gamma_c$ is the coasting Lorentz factor. In order to constrain Lorentz factor using the thermal emission information, the photosphere radius $R_{\rm ph}$ needs to be greater than $R_c$. In previous treatments \citep[e.g.][]{meszarosrees00,peer07}, $\Gamma_c$ is approximated as $\eta$ (for the regime we are interested in, i.e. $R_{\rm ph}>R_c$). We note that the fireball Lorentz factor never fully achieves $\eta$, as the fireball contains a significant amount of internal energy, especially below $R_{\rm ph}$. Numerical simulations \citep{kobayashi99} showed that acceleration does not stop abruptly, but undergoes a smooth transition around $R_c$. See also Figure \ref{fig:cartoon}. As a result, a more reasonable approximation would be that the Lorentz factor of the fireball only reaches $\Gamma_{\rm ph}$ at $R_{\rm ph}$, when the fireball becomes transparent. After discharging photons at $R_{\rm ph}$, the internal energy becomes negligibly small so that $\mu$ becomes close to the bulk Lorentz factor $\Gamma = \Gamma_{\rm ph}$, which coasts with this value afterwards. As a result, one may approximately treat the fireball dynamics as having an effective coasting Lorentz factor $\Gamma_c \sim \Gamma_{\rm ph}$ and an effective coasting radius at $R_c \sim \Gamma_{\rm ph} R_0$.

For $R_{\rm ph}>R_c$ (i.e. $\Gamma_{\rm ph} < \Gamma_{\rm ph,*}$), the observer-frame (without the $(1+z)$ correction from cosmological expansion) luminosity and temperature of the photosphere emission can be estimated as (\citealt{meszarosrees00}, but with $\eta$ replaced by $\Gamma_{\rm ph}$, and $L_{w,0}$ replaced by $L_{\rm w,ph}$)
\begin{eqnarray}
    \frac{L_{\rm ph}}{L_{\rm w,ph}} & \simeq &  \left(\frac{\Gamma_{\rm ph}}{\Gamma_{ph,*}} \right)^{8/3} = \left(\frac{R_{\rm ph}}{R_c} \right)^{-2/3}  = \left(\frac{r_{\rm ph}}{R_0} \right)^{-2/3}, \label{eq:L} \\
    \frac{T_{\rm ph}}{T_0} & \simeq &  \left(\frac{\Gamma_{\rm ph}}{\Gamma_{\rm ph,*}} \right)^{8/3} = \left(\frac{R_{\rm ph}}{R_c} \right)^{-2/3} = \left(\frac{r_{\rm ph}}{R_0} \right)^{-2/3}, \label{eq:T}
\end{eqnarray}
where $L_{\rm ph}$ is the photosphere emission luminosity (i.e. the luminosity of the thermal spectral component), $L_{\rm w,ph}$ is the kinetic luminosity of the wind at the photosphere, which is related to the total wind luminosity through $L_{\rm w,ph} = L_{w,0} (\Gamma_{\rm ph} / \eta)$, 
\begin{equation}
 r_{\rm ph} = \frac{R_{\rm ph}}{\Gamma_{\rm ph}}
\end{equation}
is the radius of the projected photosphere area for a relativistically moving fireball,
\begin{equation}
    \Gamma_{\rm ph,*} = \left( \frac{L_{\rm w,ph}  {\cal Y}\sigma_{\rm T}}{8\pi m_p c^3 R_0} \right)^{1/4} \simeq 870 \left(\frac{L_{\rm w,ph,52} {\cal Y}}{R_{0,7}}\right)^{1/4}
\end{equation}
is the critical $\Gamma_{\rm ph}$ above which $R_{\rm ph}$ becomes smaller than $R_c$ so that the method discussed here no longer applies, and
\begin{equation}
    T_0 \simeq \left( \frac{L_{w,0}}{4\pi R^2_0 \sigma_{\rm B}} \right)^{1/4} \simeq 1.9 \times 10^{10} \ {\rm K} \left(\frac{L_{w,0,52}}{R_{0,7}}\right)^{1/4}
\label{eq:T0}
\end{equation}
is the initial temperature at the central engine. Here $m_p$ is the proton mass, $c$ is the speed of light, $\sigma_{\rm T}$ is the Thomson cross section,  $\sigma_{\rm B}$ is the Stefan-Boltzmann constant, $\cal Y$ is the lepton-to-baryon number ratio, which equals unity for a pure hydrogen fireball but could be greater (for a pair-loaded fireball) or slightly smaller (for a neutron-rich fireball without pair loading) than unity. Both $L_{\rm w,ph}$ and $L_{w,0}$ are normalized to $10^{52} \ {\rm erg \ s^{-1}}$ (hereafter the convention $Q=10^n Q_n$ is adopted in cgs units). Notice that in Eq. (\ref{eq:T0}) we have neglected a coefficient of order unity, which depends on the composition of the outflow at the jet base \citep{kumarzhang15}. Other coefficients of the order unity are also neglected in our derivations below.

The observed flux of the photosphere blackbody\footnote{The photosphere spectrum is not exactly a blackbody, but does not significantly deviate from it \citep{peer12,deng14b}.} component is $F_{\rm bb}^{\rm ob} = (4\pi r_{\rm ph}^2 \sigma_{\rm B} T_{ph}^4) / (4\pi D_{\rm L}^2)$. Using Equation (\ref{eq:T}) and noticing $L_{w,0} = 4 \pi D_{\rm L}^2 F_\gamma^{\rm ob} \eta_\gamma^{-1}$ ($F_\gamma^{\rm ob}$ is the observed total gamma-ray flux), one can derive \citep{peer07}
\begin{equation}
    R_0 \simeq \frac{D_{\rm L}}{(1+z)^2} \eta_{\rm th}^{3/2}  {\cal R},
\label{eq:R0}
\end{equation}
where
\begin{equation}
    {\cal R} \equiv \left(\frac{F_{\rm bb}^{\rm ob}}{\sigma_{\rm B} T_{\rm ob}^4} \right)^{1/2} \simeq \frac{r_{\rm ph}}{D_{\rm L}} (1+z)^2,
\end{equation}
\begin{equation}
    \eta_{\rm th} \equiv \frac{\eta_\gamma F_{\rm bb}^{\rm ob}}{F_\gamma^{\rm ob}} = \frac{E_{\rm th}}{E_{\rm tot}},
\end{equation}
and $T_{\rm ob} = T_{\rm ph}/(1+z)$ is the effective temperature of the observed thermal spectrum. 

Making use of Equation (\ref{eq:L}) and noticing $L_{\rm w,ph} = 4\pi D_{\rm L}^2 F_\gamma^{\rm ob} f_\gamma^{-1}$,
where 
\begin{equation}
f_\gamma = \frac{L_\gamma}{L_{\rm w,ph}} =
\frac{\eta-\Gamma_0}{\Gamma_{\rm ph}},
\label{eq:fgam}
\end{equation}
one can further derive
\begin{eqnarray}
    \Gamma_{\rm ph} & \simeq & \left[(1+z)^2 D_{\rm L} \frac{{\cal Y} \sigma_{\rm T} F_\gamma^{\rm ob}}{2 m_p c^3 {\cal R}} \frac{f_\gamma^{1/2}}{\eta_\gamma^{3/2}} \right]^{1/4} \nonumber \\
    & = &  \left[(1+z)^2 D_{\rm L} \frac{ {\cal Y}\sigma_{\rm T} F_\gamma^{\rm ob}}{2 m_p c^3 {\cal R}} \frac{\eta^{3/2}}{(\eta-\Gamma_0) \Gamma_{\rm ph}^{1/2}} \right]^{1/4}. 
\end{eqnarray}
One can see that the parameters $\eta$ and $Y$ in Equation (4) of \cite{peer07} are replaced by $\Gamma_{\rm ph}$ and $f_\gamma^{1/2}/\eta_\gamma^{3/2}$, respectively. In the second equation, Equation (\ref{eq:fgam}) has been used. Solving for $\Gamma_{\rm ph}$, one can further derive
\begin{equation}
    \Gamma_{\rm ph} 
     =  \left[(1+z)^2 D_{\rm L} \frac{ {\cal Y}\sigma_{\rm T} F_\gamma^{\rm ob}}{2 m_p c^3 {\cal R}} \frac{\eta^{3/2}}{\eta-\Gamma_0} \right]^{2/9}.
\label{eq:Gammaph}
\end{equation}

\subsection{Afterglow constraint}\label{sec:afterglow}

For a constant density interstellar medium\footnote{We do not discuss the case of a wind medium \citep{dailu98c,meszaros98,chevalier99} in this paper. Afterglow observations suggest that the majority of GRBs, especially those with the clear deceleration signature, are consistent with having a constant density medium \citep{zhang07a,liang10}.}, one can estimate $\Gamma_0$ using the observed deceleration time $t_{\rm dec}$. The deceleration radius can be estimated with $(4\pi/3) R_{\rm dec}^3 n m_p c^2 = E_k / (\hat\gamma \Gamma_{\rm 0} \Gamma_{\rm dec})$, where $\Gamma_{\rm dec} =\Gamma_0/2$. This gives the deceleration radius $R_{\rm dec}=(3E_k/2\pi \hat\gamma \Gamma_0^2 n m_p c^2)^{1/3} \simeq (6.2 \times 10^{16} \ {\rm cm}) E_{k,52}^{1/3} \Gamma_{0,2}^{-2/3} n^{-1/3}$. The deceleration time in the observer frame can be calculated as $t_{\rm dec} = \int_0^{\rm R_{\rm dec}} (1+z) / (2 \Gamma(r)^2 c) dr \simeq 0.9 (1+z) R_{\rm dec}/\Gamma_0^2 c$. Reversely solving it, one finally gets \citep{zhang18}
\begin{eqnarray}
    \Gamma_0 & \simeq & 0.9^{3/8} \left( \frac{3E_k (1+z)^3}{2\pi \hat\gamma n m_p c^5 t_{\rm dec}^3} \right)^{1/8} \nonumber \\
    & \simeq & 170 t_{\rm dec,2}^{-3/8} \left(\frac{1+z}{2}\right)^{3/8} \left(\frac{E_{k,52}}{n}\right)^{1/8} \nonumber \\
    & = & 170 t_{\rm dec,2}^{-3/8} \left(\frac{1+z}{2}\right)^{3/8} \left(\frac{E_{\gamma,52}}{n}\right)^{1/8} \left(\frac{\Gamma_0}{\eta-\Gamma_0} \right)^{1/8}.
\label{eq:Gamma0}
\end{eqnarray}

\subsection{Dissecting fireball energy budget}

The five unknown parameters that characterize a GRB fireball, i.e. $\eta$, $\Gamma_{\rm ph}$, $\Gamma_0$, $\eta_\gamma$, and $M$ can be in principle solved with Equations (\ref{eq:Eth}), (\ref{eq:Enth}), (\ref{eq:eta_gam2}), (\ref{eq:Gammaph}) and (\ref{eq:Gamma0}), using the observed quantities $E_{\rm th}$, $E_{\rm nth}$, $E_\gamma$, $F_\gamma^{\rm ob}$, $F_{\rm bb}^{\rm ob}$, $T_{\rm ob}$,  $t_{\rm dec}$ and $z$. There are only two free parameters. One is ${\cal Y}$, which depends on the composition of the fireball (pairs, protons and neutrons), but a reasonable estimate is ${\cal Y} \sim 1$. The second parameter is the density parameter $n$, which may be further constrained via afterglow modeling \citep[e.g.][]{panaitescu01,panaitescu02}. Even if it is not constrained, the solutions only weakly depend on it. One may take a standard value $n=1 \ {\rm cm^{-3}}$ when solving the problem. 

There is no analytical solution to the problem. One can numerically solve the problem using a root-finding algorithm. From Equations (\ref{eq:Eth}) or (\ref{eq:Enth}), one can solve
\begin{eqnarray}
    M & = & \frac{E_\gamma}{(\eta-\Gamma_0) c^2}, \label{eq:M} \\
    \Gamma_{\rm ph} & = & \frac{\eta E_{\rm nth} + \Gamma_0 E_{\rm th}}{E_{\gamma}}. \label{eq:Gamph}
\end{eqnarray}

From Equation (\ref{eq:Gamma0}), one can derive
\begin{equation}
    \eta = 3.19 \left(\frac{\Gamma_0}{300}\right)^{-7} t_{\rm dec,2}^{-3} \left(\frac{1+z}{2}\right)^{3} \left(\frac{E_{\gamma,52}}{n}\right) + \Gamma_0.
\label{eq:eta}
\end{equation}
Inserting Equations (\ref{eq:Gamph}) and (\ref{eq:eta}) to Equation (\ref{eq:Gammaph}), 
$\Gamma_0$ can be then solved by assigning typical values for ${\cal Y}$ and $n$. Once $\Gamma_0$ is solved, $\eta$ can be solved from Equation (\ref{eq:eta}); $\Gamma_{\rm ph}$ and $M$ can be solved from Equations (\ref{eq:Gamph}) and (\ref{eq:M}), respectively, and $\eta_\gamma$ can be solved from Equation (\ref{eq:eta_gam2}).

\section{Examples}\label{sec:example}

\begin{figure*}[t]
\begin{center}
\includegraphics[width=2\columnwidth]{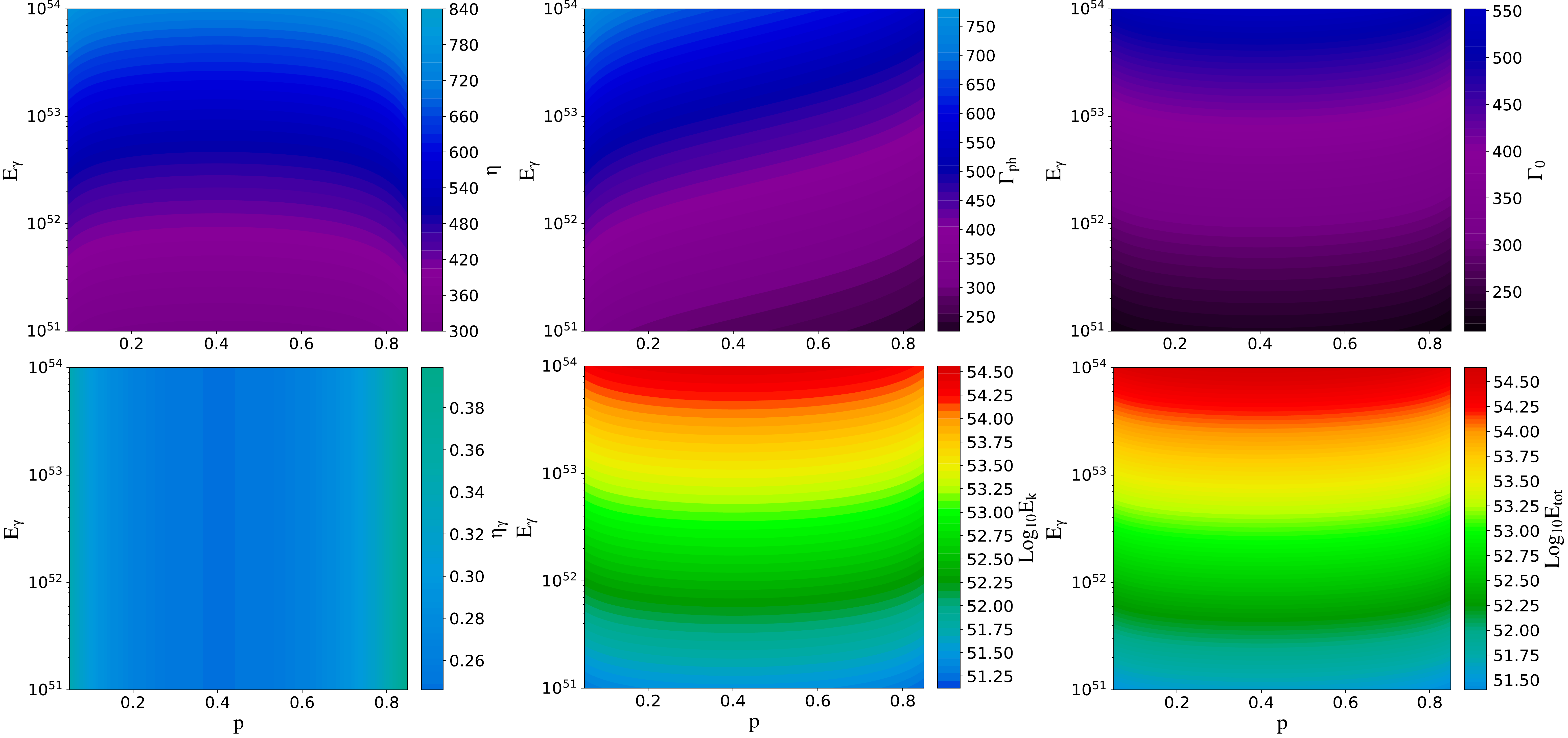}
\caption{Contour plots of  $\eta$, $\Gamma_{\rm ph}$, $\Gamma_0$, $\eta_\gamma$, $E_k$ and $E_{\rm tot}$ in the $E_\gamma - p$ plane.}
\end{center}
\label{fig:Egam-p}
\end{figure*}

\begin{figure*}[t]
\begin{center}
\includegraphics[width=2\columnwidth]{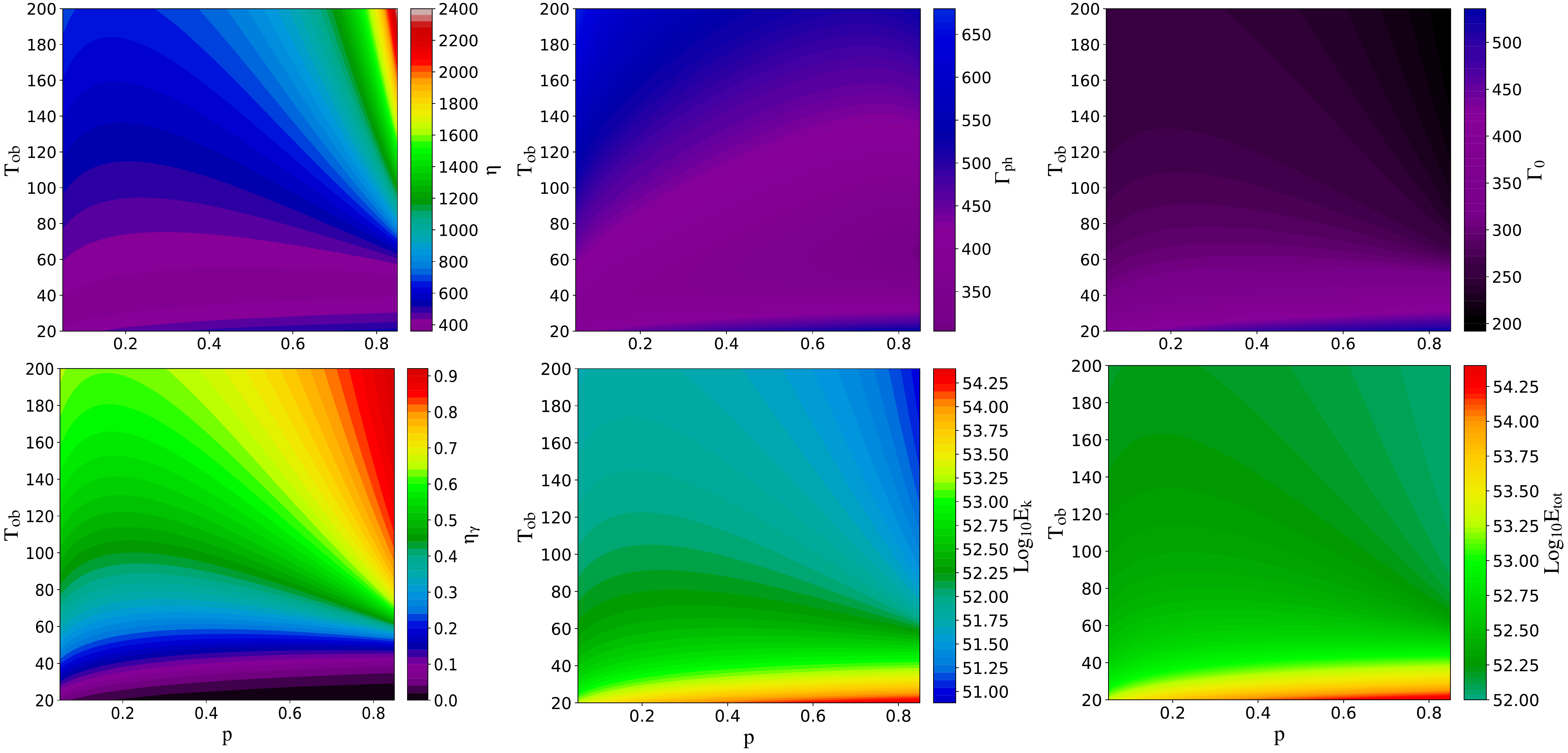}
\caption{Contour plots of  $\eta$, $\Gamma_{\rm ph}$, $\Gamma_0$, $\eta_\gamma$, $E_k$ and $E_{\rm tot}$ in the $T_{\rm ob} - p$ plane.}
\end{center}
\label{fig:T-p}
\end{figure*}

In order to perform the diagnosis proposed in this paper, a GRB needs to satisfy the following three requirements:
\begin{itemize}
    \item The burst needs to have a matter-dominated composition with a distinct thermal spectral component. One may use the contrast between the thermal and non-thermal components to estimate the magnetization parameter $\sigma_0$ at the central engine based on the hybrid-jet diagnostic method proposed by \cite{gaozhang15} (see \citealt{lil20} for a systematic analysis of the GRB data using the method). If $\sigma_0$ is close to 0, the burst would be a fireball.
    \item The burst needs to have early afterglow data that show a distinct bump that is consistent with deceleration of a fireball in a constant density medium \citep[e.g.][]{molinari07,liang10}. 
    \item The burst needs to have a measured redshift.
\end{itemize}
Few GRBs satisfy these constraints in the current database. We have gone over the currently detected GRBs from the archives, but could not find an ideal case with all three criteria satisfied. One GRB to which this method may be applied is GRB 190114C, which is studied elsewhere \citep{li21}.

Instead of performing case studies, in the following we perform calculations for some example cases and explore the dependence of the results on various parameters. For example, we consider a GRB at $z=1$ with the following observed quantities: $E_{\rm th} = 10^{53} \ {\rm erg}$, $E_{\rm nth} = 5\times 10^{52} \ {\rm erg}$, $F_\gamma^{\rm ob}=10^{-5} \ {\rm erg  s^{-1}  {cm}^{-2}}$, $F_{\rm bb}^{\rm ob} = 6\times 10^{-6} \ {\rm erg  s^{-1}  {cm}^{-2}}$, $T=100$ keV, and $t_{\rm dec} = 20$ s. According to the formalism discussed in Section \ref{sec:method}, following fireball parameters can be derived: $\eta \simeq 695$, $\Gamma_{\rm ph} \simeq 554$, $\Gamma_0 \simeq 408$, $\eta_\gamma \simeq 30.4 \%$, and $M \simeq 3.91 \times 10^{-4} M_\odot$.  

In general, the results are mainly defined by three energy values (only two are independent), i.e. $E_{\rm th}$, $E_{\rm nth}$, and $E_\gamma = E_{\rm th} + E_{\rm nth}$. This is because given a GRB duration $T_{90}$ and a redshift, the energy parameters ($E_{\rm th}$ and $E_\gamma$) can be approximately translated to the flux parameters ($F_{\rm bb}^{\rm ob}$ and $F_{\gamma}^{\rm ob}$)\footnote{$E_{\rm th}$, $E_{\rm nth}$ and $E_\gamma$ include the energies during the entire $T_{90}$ of GRB prompt emission, whereas $F_{\gamma}$ and $F_{\rm bb}$ are measured during the time intervals when the thermal emission presents. For typical GRBs, the prompt emission lightcurves show a rough fast-rise-exponential-decay behavior and the thermal emission usually appears at the most luminous peak region. For a theoretical estimation, we may calculate the flux at the peak region as $\sim 3$ times of the average flux during $T_{90}$, e.g., $F_{\gamma}^{\rm ob} \sim 3 (1+z)E_\gamma/4\pi D_{\rm L}^2 T_{90}$.}. The observed temperature $T_{\rm ob}$ is also related to $F_{\rm bb}^{\rm ob}$ through $r_{\rm ph}$. Figure \ref{fig:Egam-p} shows the contours of $\eta$, $\Gamma_{\rm ph}$, $\Gamma_0$, $\eta_\gamma$, $E_k$ and $E_{\rm tot}$ in the $E_\gamma - p$ plane, where $p \equiv E_{\rm th}/E_\gamma$ is the thermal emission fraction. The following parameters, i.e. $z=1$, $n=1 \ {\rm cm}^{-3}$, ${\cal Y}=1$, $T_{90} = 15 \ {\rm s}$, $t_{\rm dec} = 30 \ {\rm s}$, and $T_{\rm ob} = 60 \ {\rm keV}$, are adopted in the calculations. One can see that the efficiency $\eta_\gamma$ is reasonably high, between $\sim (25\% - 40\%)$ for the parameter space explored. The derived parameters $\eta$, $\Gamma_0$, $E_k$, and $E_{\rm tot}$ are all insensitive to the thermal emission fraction $p$ but positively scale with $E_\gamma$. Only the $\Gamma_{\rm ph}$ contour positively scales with both $E_\gamma$ and $p$. Fixing $E_\gamma$, $\Gamma_{\rm ph}$ decreases as $p$ increases. This is fully consistent with intuition.

Figure \ref{fig:T-p} shows the contours of $\eta$, $\Gamma_{\rm ph}$, $\Gamma_0$, $\eta_\gamma$, $E_k$ and $E_{\rm tot}$ in the $T_{\rm ob} - p$ plane. The following parameters, i.e. $z=1$, $n=1 \ {\rm cm}^{-3}$, ${\cal Y}=1$, $T_{90} = 15 \ {\rm s}$, $t_{\rm dec} = 30 \ {\rm s}$, and $E_\gamma = 10^{52} \ {\rm erg}$, are adopted for the calculations. The patterns are more complicated, which is a result of the complicated relationship between $r_{\rm ph}$ and various energy budget parameters. The bottom-left panel again shows that usually the fireball radiative efficiency $\eta_\gamma$ is high, i.e. $\sim$($20\% - 60\%$) for reasonable values of the measured blackbody temperatures and a typical observed value for $T_{\rm ob}$. Given a measured $T_{\rm ob}$, $\eta_\gamma$ increases as the thermal fraction $p$ increases to high values. This is due to the significant increase of $\eta$ in these cases.

\section{Conclusions and discussion}\label{sec:con}

We have proposed a method to dissect the energy budget of a GRB fireball making use of the constraints derived from the thermal and non-thermal emission components in the prompt emission spectrum and the deceleration bump feature in the early afterglow lightcurve of a GRB. The key point is that the blackbody spectral component observed in the prompt emission phase and the early afterglow bump are measuring the bulk Lorentz factor of the fireball at two different stages, i.e. $\Gamma_{\rm ph}$ and $\Gamma_0$, respectively. Both are lower than the initial dimensionless specific enthalpy density of the fireball $\eta$. With observational quantities such as $E_{\rm th}$, $E_{\rm nth}$, $E_\gamma$, $F_\gamma^{\rm ob}$, $F_{\rm bb}^{\rm ob}$, $T_{\rm ob}$,  $t_{\rm dec}$ and $z$, one can directly measure several crutial fireball parameters, including $\eta$, $\Gamma_{\rm ph}$, $\Gamma_0$, $\eta_\gamma$, and $M$. 

In order to apply the method, the three criteria discussed in Section \ref{sec:example} are needed. The lack of GRBs satisfying all three criteria is the combination of the rareness of fireballs and some observational selection effects. For example, the GRBs with well-studied prompt emission spectra were usually detected by {\em Fermi}, whereas those with early afterglow and redshift measurements were usually detected by {\em Swift}. On the other hand, bursts that can satisfy all three constraints may be regularly discovered by the upcoming Chinese-French GRB detector SVOM \citep{weij16}, which has the capability of obtaining both  broad-band prompt emission spectra (using ECLAIRS and GRM) and early optical afterglow lightcurves (using VT). Many of these bursts will have redshift measurements with the detection of early afterglows. The diagnosis proposed in this paper can be routinely applied to those bursts.

There are some caveats when applying the method proposed here. First, we have applied the standard fireball photosphere-internal-shock model \citep{rees94,meszarosrees00,daigne02} that invokes two distinct emission sites. Some models interpret both thermal and non-thermal emissions as arising from the photosphere region \citep[e.g.][]{vurm11,veres12b}. Our method does not apply to those models. Second, if the central engine carries significant magnetization ($\sigma_0 \gg 1$), which seems to be the case for most GRBs \citep{zhang18}, the simple method proposed here does not apply. More work is needed to extend this analysis to the case of hybrid jets following the approach of \cite{gaozhang15}. Finally, there is another channel to leak energy from the fireball, which is neutrino emission due to hadronic interactions of high-energy protons accelerated from shocks. This channel may be important for hadronic GRB models under extreme conditions \citep[e.g.][]{asano11}, but would not be important for the standard fireball model. The non-detection of neutrinos from GRBs \citep{aartsen17} suggests that the non-thermal GRB emission region is likely far from the central engine \citep{he12,zhangkumar13}, where the hadronic interaction optical depth is low. This is also consistent with the assumption that neutrino energy loss channel is unimportant. 

\acknowledgments 
We thank Peter M\'esz\'aros, Asaf Pe'er, and an anonymous referee for helpful comments.


\end{document}